\documentclass{article}

\usepackage[final]{neurips_2022_ml4ps}
\usepackage{natbib}
\usepackage{graphicx}
\usepackage{float}
\usepackage{hyperref}
\setcitestyle{square,sort,comma,numbers}

\usepackage[utf8]{inputenc} 
\usepackage[T1]{fontenc}    
\usepackage{hyperref}       
\usepackage{url}            
\usepackage{booktabs}       
\usepackage{amsfonts}       
\usepackage{nicefrac}       
\usepackage{microtype}      
\usepackage{xcolor}         

\title{Emulating cosmological multifields with generative adversarial networks}

\author{
Sambatra Andrianomena \\
  South African Radio Astronomy Observatory\\
  Cape Town, 7925\\
  Department of Physics \& Astronomy \\ University of the Western Cape\\
  Cape Town 7535\\
  \texttt{andrianomena@gmail.com} \\
  \And
  Francisco Villaescusa-Navarro \\
  Center for Computational Astrophysics \\
  Flatiron Institute \\
  New York, NY 10010 \\
  \texttt{villaescusa.francisco@gmail.com}
  \AND
  Sultan Hassan \\
  Center for Computational Astrophysics\\
  Flatiron Institute\\
  New York, NY 10010 \\ 
  Department of Astrophysical Sciences\\
  Princeton University, Peyton Hall\\
  Princeton, NJ, 08544\\
  Department of Physics \& Astronomy \\
  University of the Western Cape \\
  Cape Town 7535\\
  \texttt{shassan@flatironinstitute.org} \\ NHFP Hubble Fellow \\
}

\begin{document}

\maketitle

\begin{abstract}
  We explore the possibility of using deep learning to generate multifield images from state-of-the-art hydrodynamic simulations of the CAMELS project. We use a generative adversarial network to generate images with three different channels that represent gas density (Mgas), neutral hydrogen density (HI), and magnetic field amplitudes (B). The quality of each map in each example generated by the model looks very promising. The GAN considered in this study is able to generate maps whose mean and standard deviation of the probability density distribution of the pixels are consistent with those of the maps from the training data. The mean and standard deviation of the auto power spectra of the generated maps of each field agree well with those computed from the maps of IllustrisTNG. Moreover, the cross-correlations between fields in all instances produced by the emulator are in good agreement with those of the dataset. This implies that all three maps in each output of the generator encode the same underlying cosmology and astrophysics. 
\end{abstract}

\section{Introduction}
Multiwavelength astronomy has been identified as a priority in the 2020 decadal survey. Current and upcoming missions will survey the Universe at different wavelengths, from X-rays to radio: e.g. e-Rosita, SKA, Euclid, DESI, CMB-S4, Rubin and Roman Observatories, HIRAX, CHIME, Spherex. The data from these missions is expected to help us improve our knowledge on both galaxy formation and cosmology. In order to maximize the scientific outcome of these missions, accurate theoretical predictions are needed for different physical fields (different wavelengths). Hydrodynamic simulations are the most sophisticated tools that can be used to study and model this. Their main drawback is their computational cost, that being very large, limits the volume and resolution of these simulations \citep{Vogelsberger_review}. Being able to speed up these simulations is critical in order to perform a large variety of tasks: from parameter inference to model discrimination. Previous studies aimed at accelerating various cosmological simulations of different fields using generative models \citep{mustafa2019cosmogan, zamudio2019higan, perraudin2019cosmological, feder2020nonlinear,rodriguez2018fast, perraudin2021emulation, curtis2022cosmic}. In this work we study the possibility of using generative adversarial networks (GANs) to generate multifield images --2D maps where every channel corresponds to a different physical field-- that have the desired statistical properties for each field individually but also their cross-correlations. This is crucial in order to extract most of information from different large scale surveys. We note that while the generation of multifield images has been performed in the past \cite[see e.g.][]{tamosiunas2021investigating} this work contributes to that direction by using a more sophisticated model and a richer and more complex dataset.


\section{Methods}
\subsection{Data}
\label{data-training}

We make use of images from the CAMELS Multifield Dataset (CMD)\cite{CMD}. CMD contains hundreds of thousands of 2D maps generated from the output of state-of-the art hydrodynamic simulations of CAMELS \cite{villaescusa2021camels} for 13 different physical fields, from dark matter to magnetic fields. Every image represents a region of dimensions $25\times25\times5~(h^{-1}{\rm Mpc})^3$ at $z=0$. In this work we use images representing three different fields --gas mass density (Mgas), neutral hydrogen density (HI), and magnetic fields magnitude (B)-- generated from the IllustrisTNG LH set \cite[see][for details]{villaescusa2021camels, villaescusa2022camels1}. Every field contains 15,000 maps, each of them having $256\times256$ pixels, and the images are characterized by six numbers: two cosmological parameters ($\Omega_{\rm m}$ and $\sigma_8$) and four astrophysical parameters ($A_{\rm SN1}$, $A_{\rm SN1}$, $A_{\rm AGN1}$, $A_{\rm AGN2}$) that characterize the efficiency of supernova and AGN feedback. The multifields are produced by stacking 3 images that represent the same physical region but represent the different fields: Mgas, HI, and B. Thus, our images have $256\times256$ pixels and contain three channels.

\subsection{Model and training procedure}
\label{model}

We consider the Generative Adversarial Network (GAN) described in \cite{liu2020towards}, who introduced some novelties such that the generative model is capable of synthesizing high-resolution images while being trained with a relatively small number of examples. They prescribed a modified version of skip-layer connection \cite{he2016deep}, named \textit{Skip-Layer Excitation} (SLE), which, like the original skip-layer, allows the gradient flow to be preserved through the layers. Unlike the skip connection in \cite{he2016deep}, SLE does not require the two inputs to have the same dimension as the output results from channel-wise multiplications between the two inputs, according to \cite{liu2020towards}

\begin{equation}
    \bold{y} = \mathcal{F}(\bold{x}_{\rm low}, \bold{W}) \cdot \bold{x}_{\rm high},  
\end{equation}
where $\bold{x}_{\rm low}$ and $\bold{x}_{\rm high}$ are two inputs which are feature maps with lower and higher resolution respectively, $\mathcal{F}$ denotes a module with learnable weights $\bold{W}$ that operates on $\bold{x}_{\rm low}$. The generator \textit{G} in our model contains three SLE modules that perform channel-wise multiplications of maps at $4\times4$, $8\times8$ and $16\times16$ resolutions with maps of $64\times64$, $128\times128$,  and $256\times256$ resolutions respectively. 

\begin{figure}
  \centering
  \includegraphics[width=0.8\textwidth]{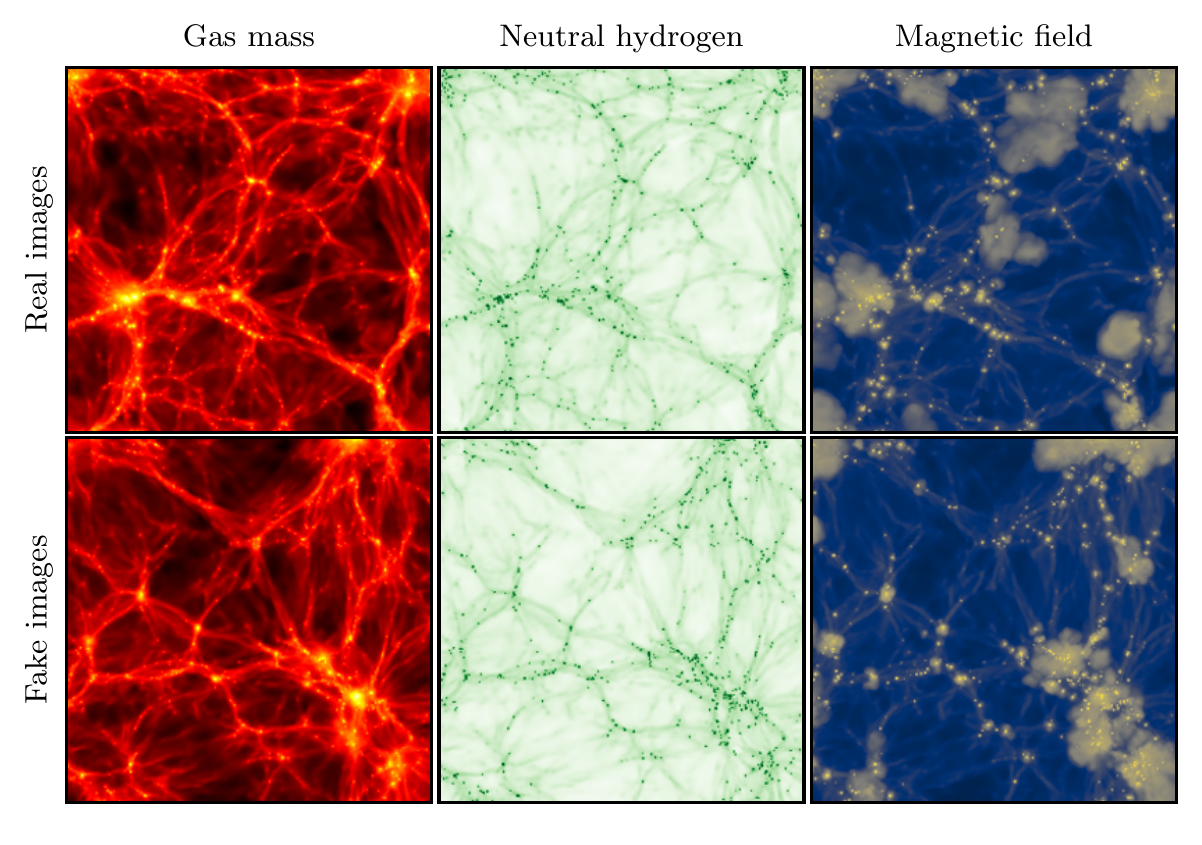}
  \caption{The top row shows images of gas mass density (left), neutral hydrogen mass density (middle), and magnetic fields strength (right) from a state-of-the hydrodynamic simulation from the CAMELS project. Note that three images represent the same region in space. The bottom row shows the same but from the our GAN model. From visual inspection we can see that the morphological features from the different fields are very well reproduced by the model.}\label{fig:real_fake_maps}
\end{figure}

Following the prescription in \cite{liu2020towards}, the discriminator \textit{D}, treated as an encoder, is trained with three decoders in order to enhance its ability to extract the salient features from the input maps. The auto-encoder, which consists of \textit{D} and the decoders, is optimized using only real images by minimizing the reconstruction loss \cite{liu2020towards}
\begin{equation}
    \mathcal{L}_{\rm recons} = \mathbb{E}_{\bold{f}\sim D(x)}[||\mathcal{G}(\bold{f}) - \mathcal{T}(x)||],
\end{equation}
where $\bold{f}$ denotes intermediate feature maps with resolutions $8\times8$ and $16\times16$ from \textit{D}, $\mathcal{G}$ is the decoding process and $\mathcal{T}$ represents some transformations on the real images, namely cropping and downsampling, on the real images $x$. For the adversarial training, we have that \cite{liu2020towards}
\begin{equation}
    \mathcal{L}_{D} = -\mathbb{E}_{x\sim p_{\rm data}(x)}[{\rm min}(0, -1 + D(x))] - \mathbb{E}_{z\sim p_{z}(z)}[{\rm min}(0, -1 - D(G(z)))] + \mathcal{L}_{\rm recons}
\end{equation}
as the loss function for \textit{D} and 
\begin{equation}
    \mathcal{L}_{G} = -\mathbb{E}_{z\sim p_{z}(z)}[D(G(z))],
\end{equation}
the loss for \textit{G}. It is worth noting that $z$, which is a vector from latent space, is drawn from a standard normal distribution. Our implementation is based on \href{https://github.com/odegeasslbc/FastGAN-pytorch}{this}. We emphasize that our model is currently not able to perform the image generation conditioned on the value of the cosmological and astrophysical parameters.

For training we set the learning rate to 0.0002 and employ the Adam optimizer. The model is trained with 10,000 multifield images for 700 epochs, which take about 12 minutes each, in batches of 8 instances on a NVIDIA GeForce GTX 1080 Ti.

\section{Results}

Fig. \ref{fig:real_fake_maps} shows the three channels from a real (top row) and generated (bottom row) multifield image. From visual inspection we can see that the morphological features associated to each field, from voids, filaments, halos, and bubbles, are generated with good precision. We now make use of several summary statistics to quantify the agreement between the statistical properties of the true and generated multifield images.


\begin{figure}
  \centering
  \includegraphics[width=0.75\textwidth]{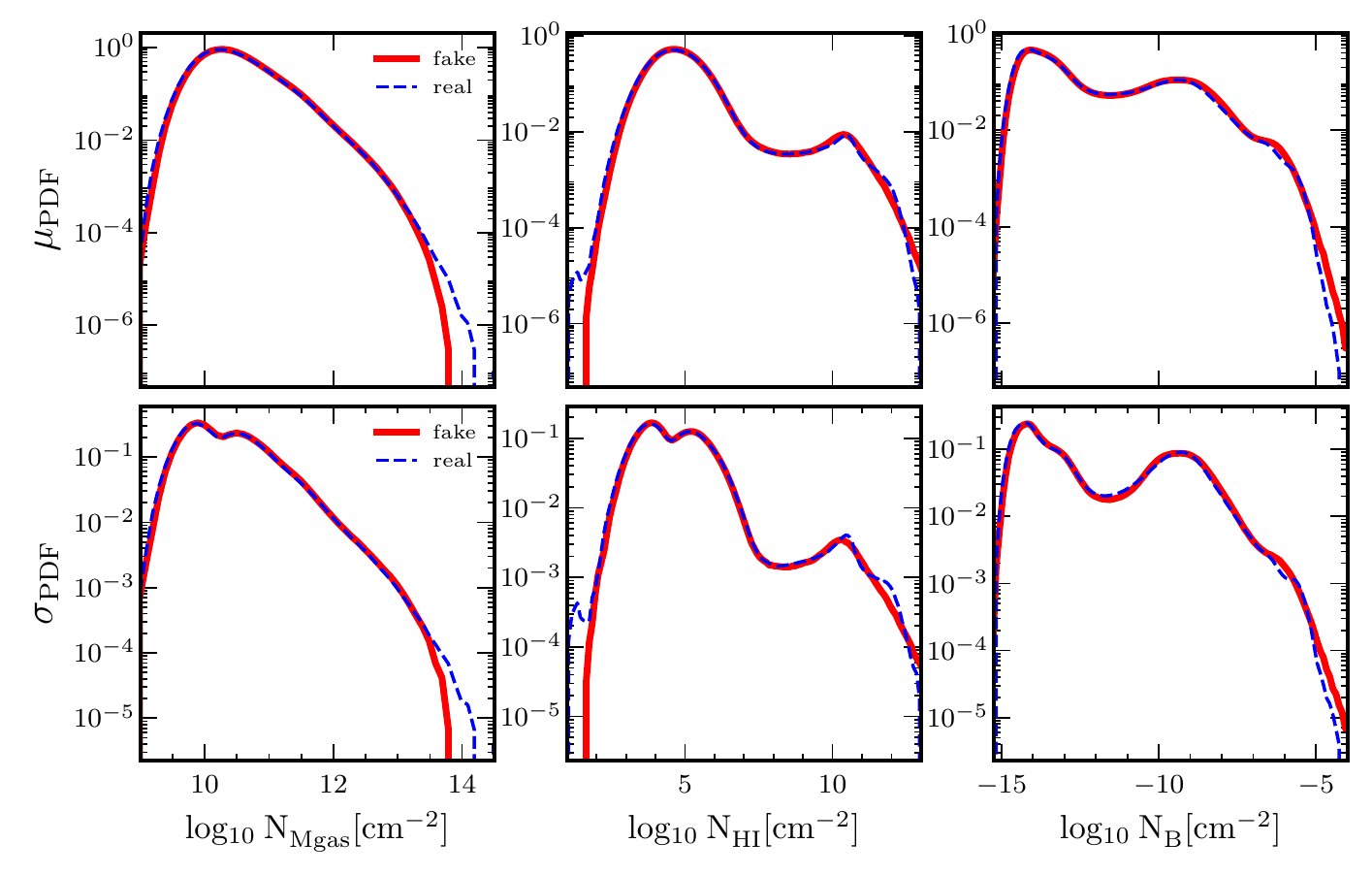}
  \caption{We have computed the probability distribution function (PDF) for 1,500 unseen real images (red) plus 1,500 generated images (blue) for each channel (field). This plot shows the mean (top) and standard deviation (bottom) of the PDF as a function pixel value for the three different fields. The agreement between the results of the true and generated images is good overall, however some statistical differences at the tails of the distributions can be noticed.}
  \label{fig:pdf_real_fake}
\end{figure}

We first consider the probability distribution function of the pixel values of the different images. We compute the $\mu_{\rm PDF}$ and standard deviation $\sigma_{\rm PDF}$ as a function of pixel value for each field using 1,500+1,500 unseen+generated multifield images. Results are shown in Fig. \ref{fig:pdf_real_fake}. We find that our GAN is able to encode the key features of each field, as suggested by the overall good agreement between the $\mu_{\rm PDF}$ of real and fake maps in each channel (see top panel in Figure\ref{fig:pdf_real_fake}). The ability of the generator to capture the variability of the pixel distribution of each field suggests that it has learned a robust lower dimension representation of the data.

\begin{figure}
  \centering
  \includegraphics[width=0.75\textwidth]{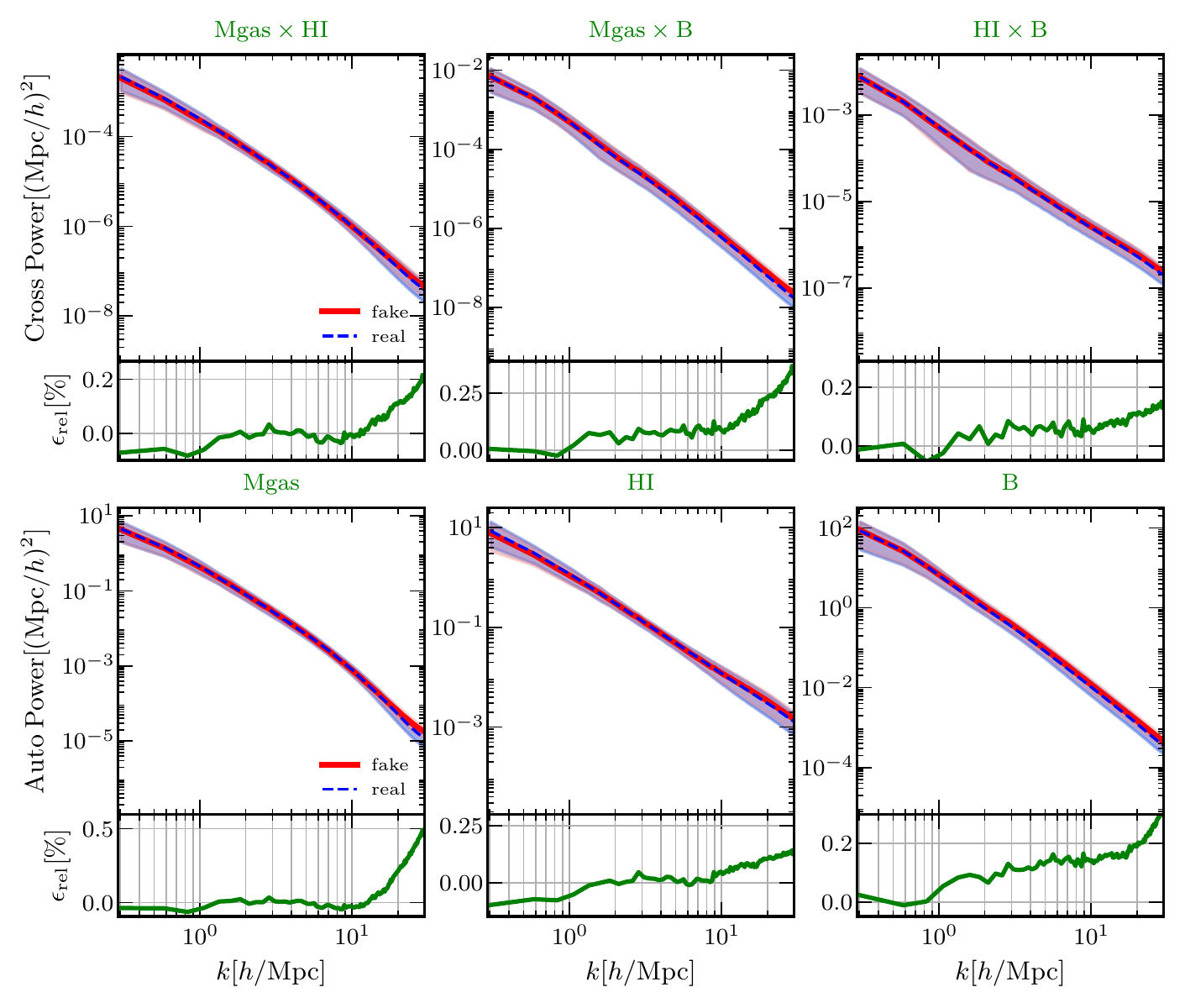}
  \caption{Same as Fig. \ref{fig:pdf_real_fake} but for the cross-power spectrum (top row) and auto-power spectrum (bottom row). Relative error between the two power spectra (fake and real), denoted by solid green, is presented at the bottom part of each panel. As can be seen, the statistical agreement between the true and generated images is relatively good for this statistics. The title of each panel displays the considered field(s).}
  \label{fig:power_spectra}
\end{figure}

Next, we quantify the agreement between clustering properties of the real and generated images by computing their auto- and cross-power spectra using {\sc Pylians} \cite{villaescusa2018pylians}. For the 1,500+1,500 real and generated images we compute their auto-power spectrum for each channel and their cross-power spectrum between channels. We show the mean and standard deviation of the results for each wavenumber in Fig. \ref{fig:power_spectra} top plot in each panel, whereas the relative difference between the two power spectra at each $k$-mode $\left(\frac{P_{\rm fake}(k)}{P_{\rm real}(k)} - 1\right)$ is shown at the bottom plot in each panel (solid green). As can be seen, although the relatively larger difference on small scales ($k > 10\>h/{\rm Mpc}$), the model is able to reproduce the overall properties of the auto- and cross-power spectra from the real images very well, including their scatter induced by cosmic variance and changes in cosmology and astrophysics.  

The above statistical tests indirectly point towards the fact that our model is not affected by mode collapse, 
an issue GANs are known to suffer from, making their training challenging. 
In order to further test this, we have created multifield images from different latent-space points, and by interpolating among them (in latent-space) we find a smooth transition between the images and their channels.
\section{Conclusions and future work}
\label{conclusion}

We have used a generative model to produce multifield images with very similar statistical properties (as quantified by pdfs and power spectra) orders of magnitude faster than the ones produced from the very computationally demanding state-of-the-art hydrodynamic simulations. The agreement in the cross-power spectra indicate the models is able to preserve Fourier phases (on large scales) across channels, and therefore represents a great tool for multiwavelength studies. We note that although our model seems able to capture the effects of cosmic variance and variations in cosmology and astrophysics, it is currently unable to condition the image generation on the value of those parameters (cosmological and astrophysical). In future work we will extend our model to be able to perform this task.



\section{Broader impact}
Multiwavelength astronomy was recommended as a priority in the recent decadal survey in 2022, given its potential to improve our understanding on galaxy formation and cosmology. We have used a generative model that is able to produce multifield images from different physical fields that is orders of magnitude faster than the method used to generate the true images. Given the good statistical properties of the generated images, they can be used for a variety of task, from pipeline design to studies involving cross-correlations. Our model is general and can be applied to other datasets, such as images of galaxy fields to X-rays and SZ maps. We believe our model, and further developments on it, has the potential to play an important role in the multiwavelength astronomy era.

\section{Acknowledgments}
SA acknowledges financial support from the South African Radio Astronomy Observatory (SARAO). FVN and SH acknowledge support provided by the Simons Foundation. SH also acknowledges support for Program number HST-HF2-51507 provided by NASA through a grant from the Space Telescope Science Institute, which is operated by the Association of Universities for Research in Astronomy, incorporated, under NASA contract NAS5-26555.

\clearpage
\bibliographystyle{unsrt85}
\bibliography{references}














\end{document}